\documentclass[twocolumn,showpacs,letterpaper]{revtex4}

\usepackage{amsfonts}
\usepackage{graphicx}
\usepackage{dcolumn}
\usepackage{amsmath}
\usepackage{amssymb}
\usepackage{epsfig}

\begin{document}

\title[Short Title]{Empirical Determination of Bang-Bang Operations}
\author{Mark S. Byrd\footnote{Present address: Harvard University, Maxwell Dworkin Laboratory, 33 Oxford Street Cambridge, Massachusetts 02138} }
\email{mbyrd@chem.utoronto.ca}
\author{Daniel A. Lidar}
\email{dlidar@chem.utoronto.ca}
\affiliation{Chemical Physics Theory Group, University of Toronto, 
80 St. George Street, Toronto, Ontario M5S 3H6, Canada }
\date{\today }

\begin{abstract}
Strong and fast "bang-bang" (BB) pulses have been recently proposed as a
means for reducing decoherence in a quantum system. So far theoretical
analysis of the BB technique relied on model Hamiltonians. Here we introduce
a method for empirically determining the set of required BB pulses, that
relies on quantum process tomography. In this manner an experimenter may
tailor his or her BB pulses to the quantum system at hand, without having to
assume a model Hamiltonian.
\end{abstract}


\maketitle


\section{Introduction}

Quantum computers hold great promise in solving certain computational
problems faster than their classical counterparts, but they are notoriously
susceptible to decoherence (deviations from unitary dynamics) and unitary
errors, the combination of which we refer to as \textquotedblleft
noise\textquotedblright . The effect of decoherence is to induce
computational errors that destroy the quantum speed-up: a decohered quantum
computer can be efficiently simulated by a classical computer \cite{Aharonov:96a}. Hence the ultimate success of quantum information processing
depends on the ability to implement error correction or avoidance
techniques. To this end, a variety of quantum error correcting codes (QECC)
and other methods have been designed. These methods all share an important
feature: they are designed to deal with \emph{specific models} of errors,
as embodied in an assumed system-bath interaction Hamiltonian. The class of
active (e.g., stabilizer) QECC
\cite{Shor:95,Steane:96a,Knill:97b,Gottesman:97}, for example, is
designed to 
correct independent errors resulting from (up to) some fixed number, $t$, of
system-bath many-body interactions; the class of passive QECC
(decoherence-free subspaces) works optimally under the assumption of
collective (i.e., fully correlated) decoherence
\cite{Zanardi:97c,Zanardi:97a,Duan:98,Lidar:PRL98,Bacon:99a} or
assumes multiple-qubit errors \cite{Lidar:00a}; dynamical
symmetrization methods assume baths with relatively 
long correlation times and weak system-bath coupling, so that decoherence
may be suppressed using fast and strong \textquotedblleft
bang-bang\textquotedblright\ (BB) pulses, introduced in \cite{Viola:98}, and
further developed in \cite{Duan:98e,Vitali:99,Zanardi:98b,Viola:99,Zanardi:99a,Zanardi:99d,Viola:99a,Viola:00a,Agarwal:99,Search:00,Vitali:01,Agarwal:01,ByrdLidar:01,Viola:01a,ByrdLidar:01a,WuLidar:01b,WuByrdLidar:02}. In spite of this impressive arsenal of methods there is a fundamental
problem in the model-specific approach in terms of its applications to
experimental quantum information processing. The problem is that in real
world applications, decoherence is often a combined effect, which arises
from a variety of sources, and does not correspond to one particular model.
It is often very difficult to identify and isolate the various sources. The
result is that the model-dependent approach for overcoming decoherence
breaks down when applied to realistic systems, since it inevitably fails to
capture all sources. In addition, current methods tend to ignore the
experimental constraints imposed by the finiteness of resources, such as the
scarcity of qubits in present-day implementations of quantum computers
(presently, fewer than 10 qubits). Of course, this criticism in no way
diminishes the importance of the model-specific approach: it is through that
approach that ground-breaking new results have been obtained which establish
the in-principle possibility of overcoming decoherence. In particular, this
work has led to the observation that fault tolerant quantum computation is
possible in the independent errors model provided the fidelity of gate
operations is above a certain threshold \cite{Aharonov:96,Preskill:97a,Knill:98,Steane:99a,Gottesman:99a}.

We focus here on the BB method and consider a paradigm that is the reverse
of the model-dependent approach to decoherence-reduction: Instead of
assuming a specific model of decoherence, designing a corresponding QECC,
and then looking for a system that might be described to a good
approximation by that model, \emph{we propose to tailor a set of BB pulses
to a system, from experimentally measured decoherence data}. We call this
approach, which we introduced first in \cite{ByrdLidar:01a},
\textquotedblleft Empirical Bang-Bang\textquotedblright. Empirical BB is a
phenomenological approach which forsakes a microscopic understanding of the
underlying decoherence processes in favor of a direct attack on the combined
effect of all sources of decoherence at once. The procedure can be
iteratively optimized using a closed-loop learning algorithm \cite{Peirce:88,Kosloff:89,Judson:92,Bardeen:97,Lloyd:97,Rabitz:00,Brif:01}. In this manner one may take into account
practical constraints imposed by the specific physical and experimental
realization.

That empirical BB is feasible in principle follows from two key facts: (i) It is
possible to experimentally measure the superoperator (i.e., the map that
propagates the density matrix) characterizing the noise in a particular
system by using Quantum Process Tomography (QPT); (ii) As we show here,
given knowledge of the superoperator it is possible to design a BB
procedure. Thus an experiment can, in principle, provide all the information
needed to design an optimized set of BB pulses.

This article is arranged as follows. In section \ref{review} we
review the basic background to quantum process tomography, and the
theory for decoupling by symmetrization. We then present, in  section
\ref{empsbb}, a derivation and discussion of several formulas
for determining the set of decoupling operations. These results are
illustrated in section \ref{examples} with a few examples. We
then indicate in section \ref{learningalgos} how the empirically
determined set of BB pulses can be optimized using a learning loop
algorithm.


\section{Review}

\label{review}

In this section we review the important components of the empirical
determination of bang-bang operations. These include quantum process
tomography (QPT), the theory of dynamical decoupling operations for a given
Hamiltonian, and its geometrical interpretation. Readers familiar with these
concepts can choose to skip ahead to section \ref{empsbb}, although the
notation introduced in this section will be used in the remainder of the
paper.


\subsection{Quantum Process Tomography}

\label{Sect:QPT}

The dynamics of an open quantum system coupled to a bath is formally
obtained from the time-ordered evolution 
\begin{equation}
U(t)=\mathcal{T}\exp (-i\int^{t}H(t^{\prime })dt^{\prime })  \label{eq:U}
\end{equation}
under the combined system-bath Hamiltonian 
\begin{eqnarray}
H &=&H_{S}\otimes {I}_{B}+{I}_{S}\otimes H_{B}+H_{SB}  \notag \\
H_{SB} &=&\sum_{\gamma }S_{\gamma }\otimes B_{\gamma },  \label{Htot}
\end{eqnarray}
where $I$ is the identity operator, $H_{S}$ is the Hamiltonian for the
system alone, $H_{B}$ is the Hamiltonian for the bath alone, $H_{SB}$ is the
system-bath interaction Hamiltonian, and the $S_{\gamma }$ and $B_{\gamma }$
are operators on the system and the bath respectively. Tracing over the bath
degrees of freedom in order to obtain the time-evolved system density
matrix: 
\begin{equation}
\rho (t)=\mathrm{Tr}_{B}[{U}(t)\left( \rho (0)\otimes \rho _{B}(0)\right) {U}
^{\dagger }(t)],  \label{eq:rhot}
\end{equation}
where $\rho (0)$ is the initial density matrix of the (open) system, $\rho
_{B}(0)$ is the initial density matrix of the bath. It can be shown that
this agrees with the most general quantum evolution consistent with the
condition of complete positivity, known as the Kraus operator sum
representation (OSR) \cite{Sudarshan:61,Kraus:83,Schumacher:96a}: 
\begin{eqnarray}
\mathcal{E}_{t}(\rho (0)) &\equiv &\rho (t)  \notag \\
&=&\sum_{\mu \nu }A_{\mu \nu }(t)\rho (0)A_{\mu \nu }^{\dagger }(t)  \notag
\\
&=&\sum_{\alpha ,\beta }\ \chi _{\alpha, \beta }(t)K_{\alpha }\rho
(0)K_{\beta }^{\dagger }.  \label{eq:chiOSR}
\end{eqnarray}
The \emph{Kraus operators} can be related to Eq.~(\ref{eq:rhot}) through 
\begin{equation}
A_{\mu \nu }(t)=\sqrt{\lambda _{\nu }}\langle \mu |U(t)|\nu \rangle
\label{eq:A}
\end{equation}
where $|\nu \rangle ,|\mu \rangle $ are eigenvectors of the the initial bath
density matrix: $\rho _{B}(0)=\sum_{\nu }\lambda _{\nu }|\nu \rangle \langle
\nu |$ \cite{Lidar:CP01}. Since $\mathrm{Tr}[\rho (t)]=1$, they satisfy the
normalization condition: $\sum_{\mu }A_{\mu \nu }^{\dagger }A_{\mu \nu }={\ I
}_{S}$. The matrix 
\begin{equation*}
\chi _{\alpha, \beta }(t)=\sum_{\mu \nu }b_{\mu \nu ;\alpha }b_{\mu \nu
;\beta }^{\ast }
\end{equation*}
is a time-dependent, Hermitian coefficient matrix defined by a
transformation of the Kraus operators to a \emph{fixed} (i.e.,
time-independent) operator basis $K_{\alpha }$: 
\begin{equation*}
A_{\mu \nu }(t)=\sum_{\alpha }b_{\mu \nu ;\alpha }(t)K_{\alpha }.
\end{equation*}
A prescription for determining the superoperator $\mathcal{E}_{t}$ from
experimental data (QPT) was given in a number of recent papers \cite
{Poyatos:97,Chuang:97c,Buzek:98}, and has very recently been applied in NMR
experiments \cite{Childs:00}. \emph{In this paper we will take QPT to mean
the determination of the coefficient matrix $\chi _{\alpha \beta }(t)$, with
respect to a given (experimentally convenient) choice of fixed basis
operators $K_{\alpha }$. }Formally, the problem is to invert the $\chi $-matrix from experimental data. Since $\chi $ is time-dependent it is clear
that one can in practice only sample it. If the decoherence process is
Markovian then it suffices to obtain the time-independent coefficient matrix 
$A$ that appears in the Lindblad equation \cite{Lindblad:76,Alicki:87}.
However, even this is a formidable problem: if the density matrix has
dimensions $N\times N$ (where for $n$ qubits $N=2^{n}$) then a simple
counting argument shows that there are at most $N^{4}-N^{2}$ independent
real parameters in $A$ and the same number, but time-dependent, in
$\chi$. Even for 
one qubit this amounts to $12$ different parameters that may have to be
measured to completely characterize the decoherence process. Fortunately, it
is well known that in practice as few as $2$ parameters may suffice, as is
the case with the $T_{1}$ and $T_{2}$ relaxation times in NMR 
\cite{Slichter:book}.  

The general idea behind QPT is to characterize the superoperator action on a
complete basis set. To see this, let the $N^{2}$ matrices $\rho_{j}$ be a
basis for the density matrix $\rho $. For example, $\rho _{j}$ could be the
set of pure states $|j\rangle \langle j^{\prime }|$, which are then fed into
the decoherence process as inputs: $\mathcal{E}(\rho _{j})=\sum_{k}\lambda
_{jk}\rho _{j}$. Using quantum state tomography \cite{Vogel:89}, one can
experimentally determine $\lambda _{jk}$, which fully specifies the
superoperator $\mathcal{E}$, since it is now possible to find the 
$\chi $-matrix: Define $\xi $ by $K_{\alpha }\rho _{j}K_{\beta }^{\dagger
}=\sum_{k}\xi _{jk}^{\alpha \beta }\rho _{k}$, where $K_{\alpha }$ are the
fixed basis Kraus operators. Then one can show that $\sum_{\alpha \beta }\xi
_{jk}^{\alpha \beta }\chi _{\alpha \beta }=\lambda _{jk}$ \cite{Chuang:97c}.
This can be thought of as a matrix equation for the vector $\chi $ and it
can be solved by computing the inverse of the matrix $\xi $. Thus, by
measuring $\lambda $ and by giving $\xi $ through a choice of the fixed
operator basis $K_{\alpha }$, finding the $\chi $-matrix has been
transformed into a linear algebra problem. In practice, we note that it may
often be difficult to prepare the full basis set $\rho _{j}$. An interesting
alternative, using entangled input states, was recently proposed in
\cite{DAriano:01}. A method that circumvents tomography altogether
(but is less general),
using quantum network ideas, was described in \cite{Ekert:02}.


\subsection{Decoupling by Symmetrization}

\label{symmdecoup}

The process of decoupling by symmetrization counteracts noise by
applying sequences of frequent and strong pulses. The time scales are
crucial: one needs to perform a complete cycle of
symmetrization operations in a time shorter than the inverse of the
high-frequency cutoff of the bath spectral density \cite{Viola:98,Duan:98e,Vitali:99}. An elegant group-theoretical treatment shows
that the applied pulses are unitary transformations forming a
finite-dimensional group, and the application of a series of pulses amounts
to an average (symmetrization) over this group \cite{Zanardi:98b,Zanardi:99d,Viola:99,Viola:99a,Viola:00a}. A geometrical
interpretation, reviewed below, can offer further insight \cite{ByrdLidar:01}. The method can also be used to perform \textquotedblleft environment
engineering\textquotedblright , in order to prepare the conditions that
allow for DFSs \cite{Zanardi:98b,Viola:00a,WuLidar:01b}, as well as in order
to eliminate leakage errors that couple encoded states with states out
of a DFS \cite{Zanardi:98b,WuByrdLidar:02}. We briefly review this theory.

A set of symmetrization operations is chosen such that they form a discrete
subgroup of the full unitary group of operations on the Hilbert space of the
system. Denote this group $\mathcal{G}$ and its elements $g_{j}$, $
j=0,1,...,|\mathcal{G}|-1$, where $|\mathcal{G}|$ is the order of the group.
The cycle time is $T_{c}=|\mathcal{G}|\Delta t$, where $|\mathcal{G}|$ is
the number of symmetrization operations, and $\Delta t$ is the time that the
system evolves freely between operations under $U_{0}$. The symmetrized
evolution is given by 
\begin{equation}
U(T_{c})=\prod_{j=0}^{|\mathcal{G}|-1}g_{j}^{\dagger }U_{0}(\Delta
t)g_{j}\equiv e^{iH_{\rm eff}T_{c}},  \notag  \label{exactevol}
\end{equation}
where the evolution under $H_{SB}+H_{B}$ has been neglected during pulse
application, i.e., during the action of the group elements $g_{j}$. $H_{\rm eff}$
denotes the resulting effective Hamiltonian. Since the approximation
requires very strong, short pulses to be implemented in a sequence, they
have been termed bang-bang (BB) operations (we will use decoupling, symmetrization, and
BB operations interchangeably). In this (BB) limit 
\begin{equation}
H\mapsto H_{\rm eff}=\frac{1}{|\mathcal{G}|}\sum_{j=0}^{|\mathcal{G}
|-1}g_{j}^{\dagger }Hg_{j}\equiv \Pi _{\mathcal{G}}(H),  \label{eq:Heff}
\end{equation}
where $H_{\rm eff}$ is the desired Hamiltonian (without noise). The map $\Pi _{
\mathcal{G}}$ is the projector into the centralizer, $Z(\mathcal{G})$,
defined as 
\begin{equation}
Z(\mathcal{G})=\{X|\;[X,g_{j}]=0,\;\forall g_{j}\in \mathcal{G}\}.  \notag
\end{equation}
It is clear that $\Pi _{\mathcal{G}}$ commutes with all $g_{j}$ so that, if
our group is generated by $\{I,H_{S},S_{\gamma }\}$, the evolution will
proceed without the operators $S_{\gamma }$ affecting the system since the
error operators will commute with the effective Hamiltonian. The \emph{control algebra} is the algebra generated by the set $\{g_{j}\}$. Even if
the symmetrization is performed under less than ideal conditions, BB can
still reduce the noise in the system \cite{Viola:98,Duan:98e}.

The main advantage offered by dynamical decoupling is that \emph{it does not
require extra qubits}. This is a very attractive feature compared to both
active and passive error-correction, one that may make dynamical decoupling
a method of choice for small-scale quantum computer implementations,
provided its stringent time-scale requirements can be met.


\subsection{Geometry of the Decoupling Method}

\label{geomBB}

In preparation for the remainder of the paper, and as an intuitive aid, we
briefly review the geometric description of BB controls developed in \cite{ByrdLidar:01}. Let us explicitly introduce $N\equiv n^{2}-1$ traceless,
Hermitian generators $\{\lambda _{i}\}_{i=1}^{N}$ of $SU(n)$. These
generators are closed under commutation and span the space of traceless
Hermitian matrices. For $SU(2)$, the Pauli matrices are commonly used; for $
SU(3)$, the Gell-Mann matrices, and for higher dimensions, one may use a
direct generalization of the Gell-Mann matrices. For dimensions that are a
power of two (and quantum computing) it is often convenient to use the Pauli
group (tensor products of Pauli matrices). The $\{\lambda _{i}\}$ satisfy
trace-orthogonality, 
\begin{equation}
\mathrm{Tr}(\lambda _{i}\lambda _{j})=M\delta _{ij},  \label{eq:tr}
\end{equation}
where $M$ is a normalization constant (often taken to be $2$ for Lie
algebras or $n$ for $n\times n$ matrices). Expanding the system operators in
terms of the $\{\lambda _{i}\}$ yields: 
\begin{equation}
K_{\gamma }=\sum_{i}a_{i\gamma }\lambda _{i}
\end{equation}
where the expansion coefficients are 
\begin{equation}
a_{i\gamma }=\frac{1}{M}\mathrm{Tr}(\lambda _{i}K_{\gamma }).  \label{eq:ai}
\end{equation}
Using this, $H_{SB}$ can be written as as follows: 
\begin{eqnarray}
H_{SB} &=& \sum_{\gamma }S_{\gamma }\otimes B_{\gamma }=\sum_{\gamma
}\sum_{i=1}^{N}a_{i\gamma }\lambda _{i}\otimes B_{\gamma } \notag
\\
&\equiv&
\sum_{\gamma }(\vec{a}_{\gamma }\cdot \vec{\lambda})\otimes B_{\gamma }.
\label{eq:Had}
\end{eqnarray}
Here $\vec{a}_{\gamma }$ and $\vec{\lambda}$ are vectors of length $N$. In
this representation, used extensively in \cite{Mahler:book}, an $n\times n$
Hamiltonian, $H$, is a vector with coordinates $\vec{a}_{\gamma }$ for each
error $\gamma $ in an $N$-dimensional vector space spanned by the $\{\lambda
_{i}\}$ as basis vectors, with ordinary vector addition and scalar
multiplication. The open system evolution is thus described by a vector (or
vector field) in the space of possible evolutions.

Now, as is well-known, there is a homomorphic mapping between the Lie groups 
$SU(2)$ and $SO(3)$ \cite{Cornwell:97}. This mapping is generalized as
follows for $SU(n)$ and a subgroup of the rotation group $SO(N)$: 
\begin{equation}
U_{k}^{\dagger }\lambda _{i}U_{k}=\sum_{j=1}^{N}R_{ij}^{(k)}\lambda _{j},
\label{eq:R}
\end{equation}
where the matrix $R^{(k)}\in SO(N)$, the adjoint representation of $SU(n)$.

The BB operation [Eq.~(\ref{eq:Heff})] may now be viewed as a weighted sum
of rotations of the (adjoint) vectors $\vec{a}_{\gamma }$. To see this,
first let 
\begin{equation}
\vec{a}_{\gamma }^{(k)}=R^{(k)}\vec{a}_{\gamma }.
\end{equation}
This represents the rotation by $R^{(k)}$ of the coordinate vector $\vec{a}
_{\gamma }$. Next average over all rotations: 
\begin{equation}
\vec{a}_{\gamma }^{\prime }=\frac{1}{|\mathcal{G}|}\sum_{k=0}^{|\mathcal{G}
|-1}\vec{a}_{\gamma }^{(k)}.
\end{equation}
Finally, note that the effective Hamiltonian, after the BB operations, can
be rewritten as: 
\begin{equation}
H_{\rm eff}=\frac{1}{|\mathcal{G}|}\sum_{k=0}^{|\mathcal{G}|-1}U_{k}^{\dagger
}HU_{k}=\sum_{\gamma }(\vec{a}_{\gamma }^{\prime }\cdot \vec{\lambda})\otimes B_{\gamma }.  \label{eq:Hgeoeff}
\end{equation}
Eq.~(\ref{eq:Hgeoeff}) [compare to Eq.~(\ref{eq:Had})] is the desired
geometric representation of BB operations. Their effect is to simply
transform, for each error $\gamma $, the coordinates $\vec{a}_{\gamma }$ to $
\vec{a}_{\gamma }^{\prime }$. It is simplest to interpret this in the case
of storage, where we seek BB operations such that $H_{\rm eff}=0$. Since the
errors can be decomposed in the linearly independent basis set indexed by $
\gamma $, each term $\vec{a}_{\gamma }^{\prime }\cdot \vec{\lambda}$ must
vanish separately. Furthermore, since the $\lambda _{i}$ are independent
this can only be satisfied if $\vec{a}_{\gamma }^{\prime }=\vec{0}$ for each 
$\gamma $. This means that 
\begin{equation}
\vec{a}_{\gamma }^{\prime }=\left( \frac{1}{|\mathcal{G}|}
\sum_{k}R^{(k)}\right) \vec{a}_{\gamma }=\vec{0},
\end{equation}
i.e., the sum of all rotations applied to the original coordinate vector $
\vec{a}_{\gamma }$ must vanish.

Similarly, to obtain a modified evolution corresponding to a target
Hamiltonian $H_{\rm eff}^t = \sum_{\gamma} (\vec{a}^t_\gamma \cdot \vec{\lambda}) \otimes B_\gamma$, we require the weighted sum of rotations applied to the
original coordinate vector to be equal to the corresponding target
coordinate vector $\vec{a}_\gamma^{t}$. I.e., for $H_{\rm eff} \neq 0$, the
following condition should be satisfied to obtain the desired evolution: 
\begin{equation}  \label{hgeff}
\vec{a}_\gamma^{\prime} = \vec{a}_\gamma^{t}
\end{equation}
This may require a combination of switching strategies for the BB pulses 
\cite{Viola:00a}.

It should be noted that the geometrical picture is an explicit
representation of a subset of the group algebra $\mathbb{C}\mathcal{G}$
using the set of traceless Hermitian matrices and the identity as the basis.
When the coefficients of the adjoint vector are real, the resulting matrix $
H_{\rm eff}$ is Hermitian. When they are complex, the resulting matrix is not
Hermitian and the evolution is not unitary, but may still be treated
empirically, as we show below.

We now turn to
showing how to find the BB pulses directly from
experimental data, i.e., given a QPT measurement of the $\chi $-matrix.


\section{Determination of Symmetrization Operators}

\label{empsbb}

Since the BB method operates at extremely fast time-scales it is useful to
consider a short-time expansion of the OSR evolution equation
(\ref{eq:chiOSR}). To do so we follow \cite{Bacon:99,Lidar:CP01},
where it was 
shown how the OSR can be rewritten to resemble the Lindblad equation \cite{Lindblad:76,Alicki:87}. Thus, the OSR can
be rewritten as 
\begin{eqnarray}
\rho (t) &=&\rho (0)-{\frac{i}{\hbar }}[S(t),\rho (0)]  \notag
\label{eq:newOSR} \\
&+&\frac{1}{2}\sum_{\alpha ,\beta =1}^{N}\chi_{\alpha, \beta }(t)\left(
[K_{\alpha },\rho (0)K_{\beta }^{\dagger }]+[K_{\alpha }\rho (0),K_{\beta
}^{\dagger }]\right) ,  \notag \\
&&
\end{eqnarray}
where $S(t)$ is the Hermitian operator defined by 
\begin{equation}
S(t)={\frac{i\hbar }{2}}\sum_{\alpha =1}^{N}\left[ \chi_{\alpha,
0}(t)K_{\alpha }-\chi_{0,\alpha }(t)K_{\alpha }^{\dagger }\right] .
\label{eq:S}
\end{equation}
Note the similarity of Eq.~(\ref{eq:newOSR}) to the Lindblad equation 
\cite{Lindblad:76,Alicki:87}. Indeed, the Lindblad Markovian semigroup master
equation can be derived from Eq.~(\ref{eq:newOSR}) via a coarse-graining
procedure \cite{Bacon:99,Lidar:CP01}, which replaces the time-dependent 
$\chi_{\alpha, \beta }$ matrix elements with their time-averages over an
interval that is longer than the bath correlation-time, and thus longer than
the BB time-scale. An important outcome of this procedure is that the
coarse-grained $S(t)$ can be interpreted as a system Hamiltonian $H_S$
plus a Lamb shift correction \cite{Bacon:99,Lidar:CP01}. While still
exact, Eq.~(\ref{eq:newOSR}) is more amenable to a short-time
expansion than the original (equivalent) form of the OSR,
Eq.~(\ref{eq:chiOSR}). 

Note that the \textquotedblleft fixed-basis\textquotedblright\ $\{K_{\alpha
}\}_{\alpha =1}^{N}$ is completely analogous to the Hermitian generators $
\{\lambda _{i}\}_{i=1}^{N}$ of $SU(n)$ used in the geometric picture
of section \ref{geomBB}. Thus, assuming a Hermitian basis $\{
K_{\alpha } \}$ we can rewrite Eq.~(\ref{eq:S}) as 
\begin{equation*}
S(t)=i\hbar \sum_{\alpha =1}^{M}\mathrm{Im}[\chi_{\alpha, 0}(t)]K_{\alpha
}=i\hbar \, \mathrm{Im}(\vec{\chi})\cdot \vec{K},
\end{equation*}
which can be interpreted as giving the \textquotedblleft
Hamiltonian\textquotedblright\ $S(t)$ as a vector with coordinates
$\mathrm{Im}[\chi_{\alpha, 0}(t)]\}$ in a space with basis vectors
$\{K_{\alpha }\}$. 

Next we give a general method for determining BB controls from empirical
data, specialize the applicability of this method somewhat, and then treat
storage, single qubit operations and computation.


\subsection{Empirical Bang-Bang Condition}

\label{EBB-cond}

Before going into a detailed and more careful analysis, we first present a
\textquotedblleft rough\textquotedblright\ version of the empirical BB\
condition. We note two key facts:\ (i) the BB method will operate only to
undo the undesired evolution due to $S(t)$; (ii) from sections \ref{Sect:QPT}
and \ref{symmdecoup} we find that, under the action of BB controls,
the $\{ K_\alpha \}$
transform as 
\begin{eqnarray}
K_{\alpha } &\overset{\mathrm{BB}}{\rightarrow }&\frac{1}{|\mathcal{G}|}
\sum_{k}U_{k}^{\dagger }K_{\alpha }U_{k}  \notag \\
&=&\frac{1}{|\mathcal{G}|}\sum_{k}\sum_{\beta =1}R_{\alpha \beta
}^{(k)}K_{\beta }=\frac{1}{|\mathcal{G}|}\sum_{k}\left( R^{(k)}\vec{K}
\right) _{\alpha }
\end{eqnarray}
Thus, given the considerations above concerning the effect of BB\ pulses and
their geometrical interpretation, we can express the BB-modified open system
evolution as 
\begin{eqnarray}
S &=&i\hbar \mathrm{Im}(\vec{\chi})\cdot \vec{K}  \notag \\
&\overset{\mathrm{BB}}{\rightarrow }&\mathrm{Im}(\vec{\chi})\cdot \frac{1}{|
\mathcal{G}|}\sum_{k}R^{(k)}\vec{K}  \notag \\
&=&i\hbar \frac{1}{|\mathcal{G}|}\sum_{k}\sum_{\alpha \beta }\mathrm{Im}[\chi
_{\alpha, 0}]R_{\alpha \beta }^{(k)}K_{\beta }  \notag \\
&=&i\hbar \mathrm{Im}(\vec{\tilde{\chi}})\cdot \vec{K}\equiv \tilde{S},
\end{eqnarray}
where the new, BB-modified \textquotedblleft Hamiltonian\textquotedblright\ $
\tilde{S}$ is described by the new, rotated coordinate vector 
\begin{equation}
\mathrm{Im}(\vec{\tilde{\chi}})=\mathrm{Im}(\vec{\chi})\cdot \frac{1}{|
\mathcal{G}|}\sum_{k}R^{(k)}.  \label{eq:chitilde}
\end{equation}
Now, let the ideal, or desired \textquotedblleft
Hamiltonian\textquotedblright\ be described by the coordinate vector $\vec{
\chi}_{w}$, i.e., 
\begin{equation}
S_{w}=i\hbar \mathrm{Im}(\vec{\chi}_{w})\cdot \vec{K}.
\end{equation}
For storage this would correspond to the null vector, but not for
computation. The goal of the empirical BB procedure is to find rotation
matrices $R^{(k)}$ such that the difference 
\begin{equation}
\tilde{S}-S_{w}=i\hbar \left[ \mathrm{Im}(\vec{\tilde{\chi}})-\mathrm{Im}(
\vec{\chi}_{w})\right] \cdot \vec{K}=0,  \label{eq:s-sw}
\end{equation}
or more generally, is minimal. This has the simple geometric interpretation
of minimization of the distance between the BB-modified vector $\mathrm{Im}(
\vec{\tilde{\chi}})$ and the desired vector $\mathrm{Im}(\vec{\chi}_{w})$. 

The \emph{input data} is $\mathrm{Im}(\vec{\chi})$ (the output of the QPT\
measurement), $\vec{\chi}_{w}$ (the desired Hamiltonian), $\vec{K}$ (the
operator basis, with respect to which $\vec{\chi}_{w}$ and $\vec{\chi}$ are
defined). This data specifies a solution to Eqs.~(\ref{eq:chitilde}), 
(\ref{eq:s-sw}) in terms of the rotation matrices $R^{(k)}$. This solution is not unique; see, e.g., 
the example in section \ref{example}.

When the $R^{(k)}$ are found, the BB pulses can be calculated 
from the transformation connecting the adjoint representation to 
its unitary group.

\emph{Thus Eqs.~(\ref{eq:chitilde}),(\ref{eq:s-sw}) can be viewed as the
essence of the empirical BB procedure}. From here on we flesh out this
first main
result.


\subsection{Qubit Noise}

\label{qstorenoise}

The development in section \ref{EBB-cond} was cavalier in its treatment of
the indices $\alpha ,\beta $ of the fixed operator basis $\{K_{\alpha }\}$.
To be more precise, consider a quantum register of $N$ qubits. We will
derive a short time expansion of Eq.~(\ref{eq:newOSR}) under the assumption
that the system-bath interaction is linear in the system operators: 
\begin{equation}
H_{SB}^{(1)}=\sum_{i=1}^{N}\vec{\sigma}_{i}\cdot \vec{B}_{i},
\label{eq:HSBlin}
\end{equation}
where $\vec{\sigma}_{i}=(\sigma _{i}^{x},\sigma _{i}^{y},\sigma _{i}^{z})$
is the vector of Pauli matrices acting on the $i^{\mathrm{th}}$ qubit, and $
\overset{\rightarrow }{B}_{i}$ is a corresponding vector of bath operators.
This assumption will be relaxed below (Section~\ref{2qubitops}) and, as
should be clear from section \ref{EBB-cond}, is not essential for our
approach, but will make the calculations below more transparent. A Taylor
expansion of the evolution operator $U(t)$ [Eq.~(\ref{eq:U})] then reveals
that as time increases, higher and higher tensor powers of the Pauli
matrices act on the qubits: 
\begin{equation}
U(t)={I}-itH_{SB}^{(1)}+\frac{(iH_{SB}^{(1)})^{2}}{2!}t^{2}+...
\label{eq:Uexpand}
\end{equation}
where for simplicity we have assumed a time-independent Hamiltonian and set $
H_{S}=H_{B}=0$. The $O(t)$ term involves only single Pauli matrices, but the 
$O(t^{2})$ terms and higher involve tensor products of Pauli matrices. To
capture this in terms of the OSR we expand the fixed basis operators $
K_{\alpha }$ as 
\begin{equation}
K_{\vec{\alpha}_{n}}\equiv \bigotimes_{i=1}^{N}\sigma _{i}^{\alpha },
\label{eq:Kbasis}
\end{equation}
where, for the $i^{\mathrm{th}}$ qubit, $\sigma _{i}^{\alpha }$, $\alpha
=0,1,2,3$ corresponds to ${I}_{i},\sigma _{i}^{x},\sigma _{i}^{y},\sigma
_{i}^{z}$ respectively. The subscript on $K_{\vec{\alpha}_{n}}$ denotes a
vector $\vec{\alpha}_{n}=(\alpha _{1},...,\alpha _{N})$ with $n$ non-zero
entries. I.e., $K_{\vec{\alpha}_{n}}$ acts non-trivially on $n$ qubits. (We also use
$\vec{\alpha}$ for a vector of arbitrary index.)
Note that we have omitted the subscript $i$ on $\alpha $ in
Eq.~(\ref{eq:Kbasis}) in order to reduce the index clutter. There exist $
M=4^{N}$ different $K_{\vec{\alpha}_{n}}$ operators with $K_{\vec{\alpha}
_{0}}={I}\otimes \cdots \otimes {I}$ being the identity on the space of all
qubits. Here we have chosen the $K$'s to be Hermitian, and trace orthogonal:
\begin{equation}
\mathrm{Tr}(K_{\vec{\alpha}_{m}}K_{\vec{\beta}_{n}})=2^{N}\delta _{\vec{
\alpha}_{m}\vec{\beta}_{n}}.
\end{equation}
Hence they are a valid basis for all $2^{N}\times 2^{N}$ matrices.

Corresponding to this expansion of the fixed-basis operators, we can rewrite
the OSR, Eq.~(\ref{eq:chiOSR}), more explicitly as 
\begin{eqnarray}
\rho (t) &=&\sum_{m,n=0}^{N}\sum_{\vec{\alpha}_{m},\vec{\beta}_{n}}\ \chi _{
\vec{\alpha}_{m},\vec{\beta}_{n}}(t)K_{\vec{\alpha}_{m}}\rho (0)K_{\vec{\beta}
_{n}}  \notag \\
&=&\chi _{\vec{\alpha}_{0},\vec{\beta}_{0}}(t)\rho (0)  \notag \\
&+&\sum_{m=1}^{N}\sum_{\vec{\alpha}_{m}}\chi _{\vec{\alpha}_{m},\vec{\beta}
_{0}}(t)K_{\vec{\alpha}_{m}}\rho (0)+\chi _{\vec{\alpha}_{m},\vec{\beta}
_{0}}^{\ast }(t)\rho (0)K_{\vec{\alpha}_{m}}^{\dagger }  \notag \\
&+&\sum_{m,n=1}^{N}\sum_{\vec{\alpha}_{m},\vec{\beta}_{n}}\ \chi _{\vec{
\alpha}_{m},\vec{\beta}_{n}}(t)K_{\vec{\alpha}_{m}}\rho (0)K_{\vec{\beta}
_{n}}.  \label{eq:rhoexpand}
\end{eqnarray}
Thus terms that contain only single Pauli matrices but not tensor products
of Pauli matrices can only come from the second sum ($\sum_{m=1}^{N}\sum_{
\vec{\alpha}_{m}}$), with $m=1$. Comparing to Eqs.~(\ref{eq:newOSR}),(\ref{eq:S}) it is clear that this sum is responsible for (part of) the
\textquotedblleft Hamiltonian\textquotedblright\ $S(t)$, whereas the third
sum generates the Lindblad-like term in Eq.~(\ref{eq:newOSR}). (This can
also be verified directly by repeating the derivation in 
\cite{Lidar:CP01,Bacon:99} using the $K_{\vec{\alpha}_{n}}$.) Hence to first
order in $t$ we find 
\begin{eqnarray}
\rho (t)\!-\!\rho (0) &\approx &\!\!-i[S(t),\rho (0)]  \notag \\
&\approx &\!t[\sum_{\vec{\alpha}_{1}}(\chi _{\vec{\alpha}_{1},0}K_{\vec{
\alpha}_{1}}-\chi _{\vec{\alpha}_{1},0}^{\ast }K_{\vec{\alpha}_{1}}^{\dagger
}),\rho (0)]  \notag \\
&=&-t[\sum_{\vec{\alpha}_{1}}\mathrm{Im}(\chi _{\vec{\alpha}_{1},0})K_{\vec{
\alpha}_{1}},\rho (0)]  \label{eq:rhosmallt}
\end{eqnarray}
where in the last line we used the hermiticity of the $K$ operators. The
term $\sum_{\vec{\alpha}_{1}}$ is a sum over all elements of the Pauli group
with one non-identity element in the tensor product. By comparing to
Eq.~(\ref{eq:Uexpand}), and recalling the expression for the Kraus
operators, Eq.~(\ref{eq:A}), it follows that this term is directly related
to bath matrix elements of $H_{SB}$, which give rise to a Lamb shift \cite{Bacon:99,Lidar:CP01}. When the system Hamiltonian is included, it appears
in the $\sum_{\vec{\alpha}_{1}}$ term as well. However, recall that we are
developing an approach that is explicitly model-\emph{in}dependent. Hence
the only quantities we will use are the QPT-measurable $\chi _{\vec{\alpha}
_{1},0}$.

We now wish to find an appropriate set of BB controls in order to eliminate
the noise on our qubits. It should be clear from the discussion we just
presented that this noise is \emph{unitary errors} (and not decoherence), since in
the short-time limit relevant for BB we only deal with the bath-induced Lamb
shift [decoherence arises from terms that are $O(t^{2})$]. As noted
above, from sections \ref{Sect:QPT} and \ref{symmdecoup} we find that,
under the action of BB controls, the $K$ transform as 
\begin{equation}
K_{\vec{\alpha}}\overset{\mathrm{BB}}{\rightarrow }\frac{1}{|\mathcal{G}|}
\sum_{k}U_{k}^{\dagger }K_{\vec{\alpha}}U_{k}.  \label{eq:Ktrans}
\end{equation}
(Here ${\vec{\alpha}}$ denotes a vector of arbitrary index.) This transformation is the basis for much of what follows.


\subsection{Qubit Storage}

\label{qstore}

For the storage of information (without computation) in qubits, we need to
preserve the density matrix under time evolution, so that $\rho (t)=\rho (0)$. Let us denote BB-modified quantities by a tilde. In this case we should
have, using Eq.~(\ref{eq:rhosmallt}), 
\begin{equation}
\lbrack \tilde{S}(t),\rho (0)]=0
\end{equation}
as the BB control objective. Since $S$ does not contain an identity
component ${
\mathchoice {\rm {1\mskip-4.5mu l}} {\rm
{1\mskip-4.5mu l}} {\rm {1\mskip-3.8mu l}} {\rm {1\mskip-4.3mu l}}}$ we
require that 
\begin{equation}
\tilde{S}(t)=0.  \label{storesol}
\end{equation}
We proceed to turn this into a condition on BB pulses.

Recall that $K_{\vec{\alpha}_{1}}$ denotes an operator with exactly one
non-identity term (one of the three Pauli matrices acting on an unspecified
qubit). There are therefore $3N$ such operators, which we now denote
explicitly as $\sigma _{i}^{\alpha }$, where $i=1,...,N$, and $\alpha =1,2,3$. Under the assumption of a linear system-bath coupling, Eq.~(\ref{eq:HSBlin}), it is clear that the BB-pulses need only involve tensor products of
single-qubit unitaries, i.e., 
\begin{equation*}
U_{k}=\bigotimes\limits_{i=1}^{N}U_{i}^{(k)}
\end{equation*}
Then Eq.~(\ref{eq:Ktrans}) becomes 
\begin{eqnarray}
\sigma _{i}^{\alpha } &\overset{\mathrm{BB}}{\rightarrow }&\frac{1}{|
\mathcal{\ G}|}\sum_{k}U_{k}^{\dagger }\sigma _{i}^{\alpha }U_{k}  \notag \\
&=&\frac{1}{|\mathcal{G}|}\sum_{k}U_{i}^{(k)\dagger }\sigma _{i}^{\alpha
}U_{i}^{(k)}.
\end{eqnarray}
At this point it is useful to again introduce real rotation matrices $R$ to
represent the BB-group:
\begin{equation}
U_{i}^{(k)\dagger }\sigma _{i}^{\alpha }U_{i}^{(k)}=\sum_{\beta=1}^{3}R_{\alpha \beta }^{i;(k)}\sigma _{i}^{\beta }.  \label{eq:RU}
\end{equation}
Here $i$ runs over qubit indices; $k\in \{0,1,...,|\mathcal{G}|-1\}$; $
R^{i;(k)}$ is in the adjoint representation of the group $SU(2)$ [i.e., $
R^{i;(k)}\in SO(3)$] acting on the $i^{\mathrm{th}}$ qubit and has matrix
elements $R_{\alpha \beta }^{i;(k)}$. Now let us consider the transformation
of $S(t)$ under the BB controls. To simplify notation let us denote 
\begin{equation}
\xi _{\alpha }^{i}\equiv \mathrm{Im}(\chi _{\alpha,0}^{i}).  \label{eq:xi}
\end{equation}
Then from Eq.~(\ref{eq:rhosmallt}): 
\begin{equation}
\frac{i}{t}S(t)\approx \sum_{i}\sum_{\alpha }\xi _{\alpha }^{i}\sigma
_{i}^{\alpha }\equiv \sum_{i}\vec{\xi}^{i}\cdot \vec{\sigma}_{i}
\label{eq:Sapprox}
\end{equation}
\begin{eqnarray}
\frac{i}{t}S(t) &\overset{\mathrm{BB}}{\rightarrow }&\sum_{i}\vec{\xi}
^{i}\cdot \frac{1}{|\mathcal{G}|}\sum_{k}U_{i}^{(k)\dagger }\vec{\sigma}
_{i}U_{i}^{(k)}  \notag \\
&=&\sum_{i}\vec{\xi}^{i}\cdot \frac{1}{|\mathcal{G}|}\sum_{k}R^{i;(k)}\cdot 
\vec{\sigma}_{i}  \notag \\
&=&\sum_{i}\vec{\tilde{\xi}}^{i}\cdot \vec{\sigma}_{i},
\end{eqnarray}
where 
\begin{equation}
\left( \frac{1}{|\mathcal{G}|}
\sum_{k}R^{i;(k)} \right) \cdot \vec{\tilde{\xi}}^{i}  =\vec{\xi}^{i}.
\end{equation}
For storage we require $\vec{\tilde{\xi}}^{i}=0$, i.e., 
\begin{equation}
\tilde{\xi}_{\beta }^{i}=\mathrm{\mathrm{Im}}(\tilde{\chi}_{\beta
,0}^{i})=0\qquad \forall \beta ,i.  
\label{eq:storage}
\end{equation}
Thus, \emph{solving for each }$i$\emph{\ the set of linear equations} 
\begin{equation}
\sum_{k}\sum_{\alpha }\mathrm{\mathrm{Im}}({\chi }_{\alpha,0}^{i})R_{\alpha
\beta }^{i;(k)}=0  
\label{eq:linset}
\end{equation}
\emph{for the rotation matrix elements }$R^{i;(k)}$, \emph{in terms of
measurable parameters }$\chi _{\alpha ,0}^{i}$\emph{(the output of a QPT
experiment), determines the BB pulses empirically}. The pulse form of the
BB controls is determined through Eq.~(\ref{eq:RU}).

Note that if $\mathrm{Im}(\chi _{\alpha ,0}^{i})\equiv 
\mathrm{Im}(\chi _{\alpha, 0})$, i.e., there is no dependence on qubit index
(collective decoherence \cite{Zanardi:97c,Lidar:PRL98}), then the same set
of rotation matrices $\{R^{(k)}\}_{k=0}^{|\mathcal{G}|-1}$ (with matrix
elements $R_{\alpha \beta }^{(k)}$) can be used for all qubits, as already
pointed out in \cite{Viola:99} in terms of unitary BB controls. It also
shows that, for complete symmetrization, one need only ensure that $
\sum_{k}R_{\alpha \beta }^{i;(k)}=0$ for all $\alpha ,\beta $, independent
of the decoherence mechanism.

Finally, note that we can rewrite Eq.~(\ref{eq:linset}) as:
\begin{equation}
\left( \frac{1}{|\mathcal{G}|}\sum_{k}R^{i;(k)}\right) \mathrm{\mathrm{Im}}(
\vec{\tilde{\chi}}^{i})=0.
\label{eq:main1}
\end{equation}
In this manner it is clear that what we are looking for is a group of
rotation matrices $\{R^{i;(k)}\}$, acting on qubit $i$, whose average $\frac{
1}{|\mathcal{G}|}\sum_{k}R^{i;(k)}$ acts to annihilate the QPT\ measurement
output vector $\mathrm{\mathrm{Im}}(\vec{\tilde{\chi}}^{i})$. This is the
geometrical interpretation of the empirical BB condition. 
Eq.~(\ref{eq:main1}) is our second main result.


\subsection{Single-Qubit Operations}

Now suppose that we are interested in quantum \emph{computation}. In this
case we must allow for single- and two-qubit operations, such that these are
not eliminated by the BB controls. In the model-dependent approach this
translates into the (sufficient) condition that the BB generators commute with the
Hamiltonian that is implementing the computation 
\cite{Zanardi:99a,Viola:99a}. Here we derive more general conditions 
from the empirical BB perspective, which have
the advantage that they can used to determine the required set of BB pulses
directly from a QPT measurement and a stipulated, wanted system Hamiltonian.

Let us consider the case of single-qubit operations first. In this case the
system Hamiltonian need only contain a single non-identity operator (Pauli
matrix) per qubit, as in Eq.~(\ref{eq:HSBlin}). Therefore the development of
the previous subsection applies. The difference, however, is that now
instead of the storage condition of Eq.~(\ref{eq:storage}) we require the
BB-modified $\chi $-matrix elements $\tilde{\chi}_{\beta ,0}^{i}$ to assume
values that correspond to a \emph{wanted} evolution (or system Hamiltonian $
S_{w}$). Let us denote the corresponding wanted (real)$\chi $-matrix
elements by $w_{\beta }^{i}$ (they can easily be calculated from a
Hamiltonian -- see below); then the empirical BB condition 
replacing Eq.~(\ref{eq:linset}) becomes:
\begin{equation}
\frac{1}{|\cal{G}|}\sum_{k}\sum_{\alpha }\mathrm{Im}(\tilde{\chi}_{\alpha ,0}^{i})R_{\alpha
\beta }^{i;(k)}=w_{\beta }^{i}.
\end{equation}
This once again has to be solved for the rotation matrices $R^{i;(k)}$, with
elements $R_{\alpha \beta }^{i;(k)}$, given the empirical data $\vec{\tilde{
\chi}}^{i}\equiv \mathrm{Im}(\tilde{\chi}_{\alpha ,0}^{i})$. This too, 
can be written in a form amenable to a geometric interpretation: 
\begin{equation}
\left( \frac{1}{|\mathcal{G}|}\sum_{k}R^{i;(k)}\right) \mathrm{Im}(\vec{
\tilde{\chi}}^{i})=\vec{w}^{i}.
\label{eq:one-qubitgates}
\end{equation}
Now the average over the rotation matrices acts to rotate the QPT\ output
vector to a desired vector for the $i^{\mathrm{th}}$ qubit,
$\vec{w}^{i}$. Eq.~(\ref{eq:one-qubitgates}) is our third main result. 


\subsection{Two-Qubit Operations}

\label{2qubitops}

In order to implement two-qubit operations we must allow for a system
Hamiltonian that contains two-body interactions. Therefore it is useful to
comment on what happens when also the system-bath Hamiltonian contains
higher order coupling, e.g., second order: 
\begin{equation}
H_{SB}^{(2)}=\sum_{j>i=1}^{N}(\vec{\sigma}_{i}\cdot G_{ij}\cdot \vec{\sigma}
_{j})\otimes B_{ij},
\end{equation}
where $G_{ij}$ is a second-rank tensor. In this case both the third and
fourth line of Eq.~(\ref{eq:rhoexpand}) contribute terms that are bilinear
in the Pauli matrices, i.e., they contribute $\sum_{\vec{\alpha}_{2}}$ and $
\sum_{\vec{\alpha}_{1},\vec{\beta}_{1}}$ respectively. It is important to
distinguish these bilinear terms from additional bilinear terms that arise
when the expansion is taken to $O(t^{2})$. The latter may arise from $
H_{SB}^{(1)}$ [Eq.~(\ref{eq:HSBlin})] and will contribute to the non-unitary, 
decohering, part of
the evolution. However, \emph{in the context of BB controls we are only
interested in the ultra-short time limit }$O(t)$\emph{, where the bilinear
terms arising from} $H_{SB}^{(2)}$\emph{only contribute to the Lamb shift}. 
Thus, the BB pulses that are appropriate for both two-qubit operations
and a second-order system-bath Hamiltonian will be elements of $SU(4)$.

As before, the quantities extracted from the QPT measurements will be the
imaginary part of the $\chi $-matrix which we abbreviate using a matrix $\xi$, 
as in Eq.~(\ref{eq:xi}). In this case the modified evolution will provide
for the possibility of two-qubit interactions. Thus, generalizing from 
Eqs.~(\ref{eq:rhosmallt}),(\ref{eq:Sapprox}):
\begin{eqnarray}
\frac{i}{t}S(t) &\approx &[\sum_{\vec{\alpha}_{2}}\mathrm{Im}(\chi _{\vec{
\alpha}_{2},0})K_{\vec{\alpha}_{2}}+\sum_{\vec{\alpha}_{1}}
\mathrm{Im}(\chi _{\vec{\alpha}_{1}})K_{\vec{\alpha}_{1}},\rho (0)]  \notag \\
&=&\sum_{ij}\vec{\sigma}_{i}\cdot \overset{\leftrightarrow }{\xi ^{ij}}\cdot 
\vec{\sigma}_{j},
\end{eqnarray}
where
\begin{equation}
(\overset{\leftrightarrow }{\xi ^{ij}})_{\alpha \beta }=\xi _{\alpha \beta
}^{ij}\equiv \mathrm{Im}(\chi _{\alpha \beta ,00}^{ij})+\mathrm{Im}(\chi
_{\alpha 0,0 0}^{ij})  \label{compcoeffs}
\end{equation}
is a $4\times 4$ matrix of coefficients. Under the action of the set of BB
controls, 
\begin{eqnarray}
\frac{i}{t}S(t) &\overset{\mathrm{BB}}{\rightarrow }&\frac{1}{|\mathcal{G}|}
\sum_{k}\sum_{ij}U_{ij}^{(k)\dagger }\left( \vec{\sigma}_{i}\cdot \overset{
\leftrightarrow }{\xi ^{ij}}\cdot \vec{\sigma}_{j}\right) U_{ij}^{(k)} 
\notag \\
&=&\sum_{ij}\vec{\sigma}_{i}\cdot \overset{\leftrightarrow }{\tilde{\xi}^{ij}
}\cdot \vec{\sigma}_{j},
\end{eqnarray}
where 
\begin{equation}
\overset{\leftrightarrow }{\tilde{\xi}^{ij}}= \frac{1}{|\cal{G}|} \sum_{k} R^{ij;(k)} \cdot 
\overset{\leftrightarrow }{\xi ^{ij}},
\end{equation}
and the rotation matrices $R\in SO(15)$ are defined through:
\begin{equation}
U_{ij}^{(k)\dagger }\left( \sigma _{i}^{\alpha }\otimes \sigma _{j}^{\beta
}\right) U_{ij}^{(k)}=\sum_{\gamma \delta }R_{\alpha \beta ,\gamma \delta
}^{ij;(k)}\left( \sigma _{i}^{\gamma }\otimes \sigma _{j}^{\delta }\right) .
\label{eq:RU-2qubit}
\end{equation}

Again, let us describe the target, or wanted, evolution by the $\chi$-matrix 
$w$. In analogy with Eq.~(\ref{eq:RU}) the matrix $R$ is in the adjoint 
representation of the group and thus can be viewed as a rotation in the 
vector space of Hermitian matrices.  In this case, only a subgroup of 
the rotation group $SO(15)$ is represented by the adjoint action. (This 
is true for all $SU(n)$, $n\geq 3$). 
The expression analogous to Eq.~(\ref{eq:one-qubitgates})
becomes
\begin{equation}
\frac{1}{|\cal{G}|} \sum_{k} R^{ij;(k)} \cdot \overset{\leftrightarrow }{\xi ^{ij}} =
\overset{ \leftrightarrow }{w}_{ij},
\label{eq:2qubitcomp}
\end{equation}
or using explicit index notation, 
\begin{equation}
\frac{1}{|\cal{G}|} \sum_{k}\sum_{\gamma \delta }\xi _{\gamma \delta }^{ij}R_{\gamma \delta
,\alpha \beta }^{ij;(k)}=w_{\alpha \beta }^{ij}.
\label{eq:main4}
\end{equation}
Thus the two qubit case involves solving for the $225$ elements of each of
the rotation matrices $R^{ij;(k)}$, given the QPT\ data $\overset{
\leftrightarrow }{\xi ^{ij}}$ and the desired Hamiltonian $\overset{
\leftrightarrow }{w}_{ij}$. After the rotation matrices are found, one
obtains the BB\ pulses by inverting Eq.~(\ref{eq:RU-2qubit}) for the $
U_{ij}^{(k)}$. While this seems like a daunting task in general, it should
be numerically tractable, and is illustrated for a simple example in section 
\ref{example2} below. Eq.~(\ref{eq:2qubitcomp}) is our fourth main result.


\subsection{Generalization to Encoded Qubits}

Before moving on to examples, we generalize 
the empirical BB\ condition to encoded qubits, such as arise in the
theory of quantum error correcting codes (QECC) \cite
{Shor:95,Steane:96a,Knill:97b,Gottesman:97} and decoherence-free subspaces
(DFS) \cite{Zanardi:97c,Zanardi:97a,Duan:98,Lidar:PRL98,Lidar:00a,Bacon:99a}. 
In both cases it is highly desirable to let the experiment determine a
\textquotedblleft tailored encoding\textquotedblright , since the experiment
knows the decoherence processes that govern the system we wish to protect
better than any model one can design! Furthermore, combining QECC and DFS
with the BB method has proven to be a powerful tool 
\cite{ByrdLidar:01a,WuLidar:01b,WuByrdLidar:02,LidarWuBlais:02,Viola:01a}. Now,
both QECC and DFS can be described in terms of a \emph{stabilizer} 
\cite{Gottesman:97,Bacon:99a}. A stabilizer group for a set of codewords, 
i.e., a code space, is a subgroup (of the Pauli group for QECC, and of the 
of the group of all unitary transformations for DFS) that leaves the 
code space invariant. A code (whether QECC or DFS) can be completely 
specified in terms of its stabilizer \cite{Kempe:00}.

Let $\mathcal{S}$ be the vector space (group algebra) generated by real 
linear combinations of the set of generators of the stabilizer group. Any 
member of $\mathcal{S}$ will leave the code space invariant. Thus the most 
general \textquotedblleft error\textquotedblright\ we can allow in the 
outcome of a BB procedure, when compared to a given wanted Hamiltonian, is 
so that the error is in $\mathcal{S}$. Then the outcome of this ``erroneous'' 
evolution will be correct up to an overall phase. As before, let 
$\mathrm{Im}(\vec{\chi}_{w})$ be the coordinates of the 
vector corresponding to the desired
Hamiltonian evolution and $\mathrm{Im}(\vec{\tilde{\chi}})$ the actual
vector after BB operations. Formally, the condition is: 
\begin{equation}
\tilde{S}-S_{w}=i\hbar \left[ \mathrm{Im}(\vec{\tilde{\chi}})-\mathrm{Im}(
\vec{\chi}_{w})\right] \cdot \vec{K}\in \mathcal{S},  \label{eq:encoded}
\end{equation}
which should be compared to Eq.~(\ref{eq:s-sw}). This equation may be
interpreted in one of two ways. First, given an encoding and a wanted
Hamiltonian $S_{w}$, it can be solved for the BB operations that are needed
for the suppression of errors on the code subspace. Second, given a
physically implementable set of BB operations, it can be solved for a
compatible code (by finding the stabilizer). Abstractly, this procedure may
be seen as a projection of the open system evolution onto an evolution which
is in the stabilizer of the code space. The geometric projection operation
completely reduces to the group-theoretical projection onto the
commutant given in \cite{Zanardi:98b} in the case that the set of $R^{(k)}$, 
form a discrete group \cite{ByrdLidar:01}. 
The emphasis here is a 
geometric picture of the empirical operations projecting onto the stabilizer
group of the code. Note that quite generally, Eq.~(\ref{eq:encoded}) \textit{
gives an empirical means of identifying a subspace encoding such that the BB
operations drive the evolution into a subspace which does not affect the
encoded states.}  This implies a general, empirical means for the 
\emph{creation of a DFS} \cite{WuLidar:01b}.  

If the BB procedure is imperfect there will be an error component
remaining. The error vector $\vec{E}$ is given by
the difference between the BB-modified and wanted Hamiltonians in 
the $(n^{2}-1)$-dimensional vector space where our geometric
picture holds: 
\begin{equation}
\vec{E}=\mathrm{Im}(\tilde{\vec{\chi}})-\mathrm{Im}(\vec{\chi}_{w}).
\label{errorvec}
\end{equation}
The vector $\vec{E}$ gives the magnitude and direction of the error 
(\textit{i.e.}, the basis elements $\lambda _{i}$ give the type of error, 
\textit{e.g.}, bit-flip and/or phase-flip, etc.).  The corresponding 
scalar quantity is 
\begin{equation}
d(\mathrm{Im}\tilde{\vec{\chi}},\mathrm{Im}(\vec{\chi}_{w}) = 
 [\mathrm{Tr}(\mathrm{Im}(\tilde{\vec{\chi}})-
 \mathrm{Im}(\vec{\chi}_{w}))^2]^{1/2}.
\end{equation}

\begin{center}
\begin{figure}[th]
\unitlength1mm 
\begin{picture}(110,55)
\put(10,0){
\centering
\resizebox{7cm}{!}{\includegraphics{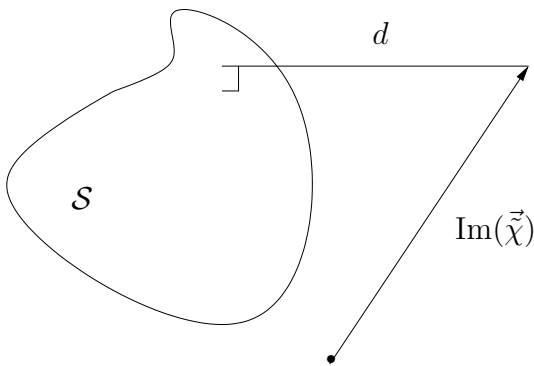}}}
\put(20,21){{\large{${\cal S}$}}}
\put(60,43){{\large{$d$}}}
\put(54.5,1){\circle*{1}}
\put(71,17){{\large{$\mathrm{Im}(\vec{\tilde{\chi}})$}}}
\end{picture}
\par
\caption{Visualization of the error $d$ that remains after the
application of BB pulses. $\mathrm{Im}(\vec{\tilde{\chi}})$ is the BB-modified
coordinate vector of the evolution, and $d$ measures the distance to
the closest element of the algebra of the stabilizer group of the code.}
\label{encodederror}
\end{figure}
\end{center}

The error (\ref{errorvec}) can be generalized to 
\begin{equation}
d(\mathcal{S},\mathrm{Im}(\vec{\tilde{\chi}})\,) 
	= \underset{\vec{B}\in \mathcal{S}}{\min }(
	  \vec{B}-\mathrm{Im}(\vec{\tilde{\chi}})\,),
\label{errorvec2}
\end{equation}
and likewise, the corresponding scalar quantity is 
\begin{equation}
d(\mathcal{S},\mathrm{Im}(\vec{\tilde{\chi}})\,) 
	= \underset{\vec{B}\in \mathcal{S}}{\min }[
	\mathrm{Tr}(\vec{B}-\mathrm{Im}(\vec{\tilde{\chi}})\,)^{2}]^{1/2},
\end{equation}
which can be visualized as in Figure~\ref{encodederror}. 

Similar conclusions were presented for the unencoded case in
\cite{ByrdLidar:01}.


\section{Examples}
\label{examples}

In this section we study a couple of simple examples that illustrate the
formalism developed above. 


\subsection{One-Qubit Example: Storing a Qubit in the Presence of Pure
Dephasing}

\label{example}

Let us consider a simple model: a phase-flip error (pure dephasing) on a
single qubit. To first order, this gives a density matrix of the form 
\begin{equation}
\rho_{s}^{\prime }\approx \rho _{s}+\left( \frac{igt}{2}\right) [\rho
_{s},\sigma _{z}],
\label{eq:exerror}
\end{equation}
where the prime indicates the density matrix for the qubit after the
interaction with a bath. The coupling constant $g$ is a measure of the strength of
the interaction. The bath time-scale is the inverse of the bath
high-frequency cutoff, which is a separate parameter. Suppose that we
wish to find a set of BB pulses that \emph{store} this qubit.

The first step in the empirical BB procedure is to measure the
superoperator using QPT. Here we would discover that the interaction causes a phase-flip error
which corresponds to $K\propto \sigma _{z}$. I.e., a measurement of the $
\chi $-matrix would yield [by comparison of Eq.~(\ref{eq:exerror}) to
Eq.~(\ref{eq:rhosmallt})]: $\{
\mathrm{Im}(\chi_{\alpha ,0}^{1})\}_{\alpha =x,y,z}=\{0,0,-g/2\}$.

The next step is to find the optimal set of BB operations. This we can do by
solving Eq.~(\ref{eq:main1}) for the rotation matrices with the measured $
\chi $-matrix. This yields 
\begin{equation}
-\frac{g}{2}\left( \sum_{k=0}^{|\mathcal{G}|-1}R_{3\beta }^{(k)}\right)
=0\;\;\Rightarrow \;\;\;\sum_{k=0}^{|\mathcal{G}|-1}R_{3\beta }^{(k)}=0.
\end{equation}
In accordance with the BB operations forming a discrete subgroup, $k=0$
corresponds to the identity. For $n=1$, since $R_{31}=0$ for the identity
rotation, 
\begin{equation}
0+\sum_{k=1}^{|\mathcal{G}|-1}R_{3\beta }^{(k)}=0\;\;\;\mathrm{and}
\;\;\;1+\sum_{k=1}^{|\mathcal{G}|-1}R_{33}^{(k)}=0,
\end{equation}
where $\beta =1,2$.

The best set of BB operations is the set that accomplishes the task 
at hand and has
the fewest elements $|\mathcal{G}|$. We now find a set with 
$|{\cal{G}}|=2$ (corresponding to a parity kick solution 
\cite{Viola:98,Vitali:99}).
I.e, we seek a rotation matrix
\begin{equation*}
R^{(1)}=\left( 
\begin{array}{ccc}
&  & 0 \\ 
&  & 0 \\ 
0 & 0 & -1
\end{array}
\right) ,
\end{equation*}
whose unspecified elements are arbitrary in as far as that they are not
determined by the QPT data. To transform from the rotation matrices back to
the BB\ pulses [i.e., from $SO(3)$ back to $SU(2)$], we use the general
result:
\begin{eqnarray}
\left( R\cdot \vec{\sigma}\right) _{\alpha } &=&e^{i\hat{n}\cdot \vec{\sigma}
\theta }\sigma_{\alpha }e^{-i\hat{n}\cdot \vec{\sigma}\theta }  \notag
\label{veceq} \\
&=&\sigma_{\alpha }\cos (2\theta )+2n_{\alpha }(\hat{n}\cdot 
\vec{\sigma})\sin ^{2}(\theta )  \notag \\
&&-(\hat{n}\times \vec{\sigma})_{\alpha }\sin (2\theta )
\end{eqnarray}
Here $R\in SO(3)$, $\hat{n}$ is a unit vector along the axis in 
$\mathbb{R}^{3}$ about which a rotation through an angle $\theta $ 
is performed (these $4$ parameters parametrize the $SO(3)$ 
rotation matrices). Let $\alpha =3$,
then we know from the form of $R^{(1)}$ that: 
\begin{equation}
-\sigma _{z}=\sigma _{z}\cos (2\theta )+2n_{3}(\hat{n}\cdot \vec{\sigma})\sin ^{2}(\theta )-(\hat{n}\times \vec{\sigma})_{z}\sin (2\theta )  \notag
\end{equation}
It is simple to check that (mod$2\pi $) the unique solution to this
equation is: $\theta =\pm \pi /2$, $n_{3}=0$. This implies $U=e^{\pm i\hat{n}
\cdot \vec{\sigma}\pi /2}$, with $\hat{n}=(n_{1},n_{2},0)$, but otherwise
arbitrary. The BB\ pulse thus must correspond to a rotation in the $x-y$
plane on the Bloch sphere, which is the expected result as the error was
along the $z$ axis.

It is likely that in a real experiment pure dephasing will not be the only
source of decoherence. Let us consider a situation where this was the
dominant source, so that our QPT\ measurement that yielded $\{\mathrm{Im}
(\chi _{\alpha ;0}^{1})\}_{\alpha =x,y,z}=\{0,0,-g/2\}$ actually contained
an $x$-component as well, which was too small to be noticed while the
dephasing process was present, e.g., because the two errors may well have
different characteristic time scales. Suppose that we perform another QPT
measurement while applying the BB pulses found above (that eliminated dephasing) and find a residual error of the $\sigma _{x}$ (bit-flip) type.
This is an instance of a learning loop, which we discuss in Section 
\ref{learningalgos} below.

In this case, consider the total Hamiltonian 
\begin{equation}
H=\frac{g^{\prime }}{2}\sigma _{x}\otimes ({I}+\sigma _{x}).
\end{equation}
Proceeding in exactly the same manner as before we determine the required BB
operations. We find that we need to implement $U=e^{\pm \hat{n}\cdot \vec{
\sigma}\pi /2}$, where now $\hat{n}=(0,n_{2},n_{3})$. Combining this and the
condition $\hat{n}=(n_{1},n_{2},0)$, we find that we need to use $\hat{n}
=(0,n_{2},0)$. Thus bit and phase flips can be corrected using the
corresponding single BB operation, which is determined empirically from an
experiment with a learning loop process. This is an optimal set since it
will eliminate both errors with only one (non-identity) BB pulse per cycle.


\subsection{Two-Qubit Example:\ Computation Using the Heisenberg Interaction
in the Presence of Independent Dephasing}

\label{example2}

As indicated above, the problem in the two-qubit case can be quite involved
since in general it requires finding the elements of rotation matrices in $
SO(15)$. To illustrate the formalism we consider a simple example. Suppose
we wish to implement a Heisenberg exchange interaction $J\vec{\sigma _{1}}
\cdot \vec{\sigma _{2}}$ between the two qubits (Heisenberg exchange is
important in a number of promising solid state proposals, and is an
interaction that is all by itself universal for QC; see, e.g., \cite
{LidarWu:01}, and references therein). Then the wanted, Heisenberg
interaction is determined from
\begin{equation*}
H_{\mathrm{Heis}}=J\vec{\sigma _{1}}\cdot \vec{\sigma _{2}}
=\sum_{ij}\sum_{\alpha \beta }{\sigma }_{i}^{\alpha }\cdot w_{\alpha \beta
}^{ij}\cdot {\sigma }_{j}^{\beta },
\end{equation*}
so that it is described by the matrix 
\begin{equation}
\overset{\leftrightarrow }{w}_{12}=J\left( 
\begin{array}{ccc}
1 & 0 & 0 \\ 
0 & 1 & 0 \\ 
0 & 0 & 1
\end{array}
\right) .
\end{equation}
Further, suppose that our QPT\ measurements suggest that the source of
decoherence in the experiment is independent dephasing on the two qubits.
This will be detected through QPT by producing the following: 
\begin{equation}
\frac{i}{t}S(t)\approx g_{1}\sigma _{1}^{z}+g_{2}\sigma
_{2}^{z}=\sum_{ij}\sum_{\alpha \beta }{\sigma }_{i}^{\alpha }\cdot {\xi }
_{\alpha \beta }^{ij}\cdot {\sigma }_{j}^{\beta }.
\end{equation}
Independent dephasing will thus be described by the matrix elements 
\begin{equation}
\xi _{3,0}^{12}=g_{1},\;\;\;\xi _{0,3}^{12}=g_{2}.
\end{equation}
To find the set of BB\ pulses we would now need to solve Eq.~(\ref{eq:main4}) for the rotation matrix elements, and then determine the
corresponding $SU(4)$ transformations, in a manner analogous to what we did
above in the single-qubit example. As noted above 
(see also \cite{ByrdLidar:01}) solving these equations is not, 
in general, trivial. In this simple example, however, a set of 
BB controls can be found noting the trace orthogonality of the 
two algebraic basis elements \cite{ByrdLidar:01}, those corresponding 
to the exchange and those corresponding to the errors. 
Rather than going through a full derivation,
we present the solution. To remove the independent dephasing through a
parity-kick procedure, without affecting the Heisenberg exchange
interaction, it is possible to use independent qubit interactions which form
the following pulse 
\begin{equation}
U\equiv U_{1}U_{2}=\exp (-i(\sigma _{1}^{x}+\sigma _{2}^{x})\pi /2)=-\sigma
_{1}^{x}\sigma _{2}^{x}.
\end{equation}
By direct calculation one can show that 
\begin{equation}
\lbrack U,H_{\mathrm{Heis}}]=0\;\;\;\mbox{and}\;\;\;\{U,S\}=0.
\end{equation}
The first commutation relation ensures that the parity-kick pulse can be
applied during computation with $H_{\mathrm{Heis}}$, while the second
(anti-)commutation relation is the parity-kick condition \cite
{Viola:98,Vitali:99,ByrdLidar:01}. Thus the desired evolution is achieved.
The pulse $U$ is certainly not unique, and a general solution of 
Eq.~(\ref{eq:2qubitcomp}) would yield a variety of other possible pulses.


\section{Optimization Algorithms}
\label{learningalgos}

As indicated in the single-qubit example discussed in the previous section,
the empirical BB procedure can benefit from the incorporation of an off-line learning
loop, that acts as an optimization algorithm for the BB pulses. Such
learning loops have proven very successful, e.g., in quantum chemical
applications, where they are typically used to optimize the yield of a
chemical reaction, steer a system towards a desired state, or perform a
cooling task \cite{Peirce:88,Kosloff:89,Judson:92,Bardeen:97,Rabitz:00,Brif:01}. Roughly,
the idea is to guide a quantum system toward a desired goal by letting a
learning algorithm optimize a classical control field (e.g., a laser pulse).
An initial field is guessed and applied to the quantum system. The output is
measured and input into a search algorithm 
(e.g., a genetic algorithm \cite{Judson:92,Goldberg:book}), 
which tries to optimize the field in order to
get closer to the desired goal. The experiment is then repeated with the new
field, and the process is repeated until it converges to the desired goal to
within a prescribed tolerance.


\subsection{Variational Optimization}

\label{optimization}

We first present an outline of a variational optimization procedure, which
can in principle be used to tailor our BB\ pulses. Our presentation follows
the standard approach in the quantum control literature, e.g., \cite{Peirce:88,Kosloff:89}. The general control problem can be stated as follows. We seek a
system Hamiltonian, $H_{c}$, which modifies a given (total, system-bath)
Hamiltonian $H$, so as to produce the desired effective Hamiltonian 
\begin{equation}
\tilde{H}=H+H_{c}.  \label{controlH}
\end{equation}
The control Hamiltonian $H_{c}$ may be composed of several possible terms, 
\begin{equation}
H_{c}=\sum_{i}u_{i}(t)H_{c}^{i},
\end{equation}
where the $u_{i}(t)$ are usually pulses in QC, analogous to the control
fields in NMR and quantum optical systems. I.e., the $u_{i}(t)$ are control
fields that may be turned on and off as desired. The unitary evolution will
proceed as usual according to $U(t)=\mathcal{T}\exp (-i\int^{t}\tilde{H}
(t^{\prime })dt^{\prime })$ with $H$, $H_{c}$ acting simultaneously. Thus
controllability is determined by the group space that one is able to
generate by the exponentiated vector fields $\tilde{H}$ \cite{Jurdjevic:72}. For the robust storage of a qubit using BB controls we
require the elimination of the interaction Hamiltonian $H$. This would
correspond to having $U\approx I$. We also wish to use as few BB operations
as possible due to the time constraints. Thus we seek to minimize the
difference between the BB-modified Hamiltonian $\tilde{S}$, and the desired
Hamiltonian $S_{w}$: 
\begin{equation}
\Delta S\equiv \tilde{S}-S_{w},  \label{eq:opcond}
\end{equation}
where $\tilde{S}$ and $S_{w}$ are the appropriate modifications to Eq. (\ref{eq:S}). E.g., for storage we would want $S_{w}=0$. One may now consider the
standard controllability problem in terms of a desired state of the system,
to be reached from some initial state $|a_{0}\rangle \equiv |a(t=0)\rangle $: 
\begin{equation}
i|\dot{a}(t)\rangle =\tilde{H}|a(t)\rangle
\end{equation}
(formally we should have included bath states as well, but we omit these for
notational convenience). $\tilde{H}=\tilde{H}(u_{i}(t),t)$ and we
may formally write the solution as 
\begin{eqnarray}
|a(t)\rangle &=&U(t,t_{0})|a_{0}\rangle  \notag \\
&=&\left[ \mathcal{T}\exp \left\{ \int_{t_{0}}^{t}\tilde{H}(u_{i}(\tau),\tau )d\tau \right\} \right] |a_{0}\rangle ,
\end{eqnarray}
where $\mathcal{T}$ is the time-ordering operator (the $H(u_{i}(\tau ),\tau) $ do not necessarily commute). This can be seen as essentially a
Heisenberg picture control problem \cite{Viola:99} and one can thus
eliminate the direct inclusion of the state itself. The short-time
approximation enables us to remove the time-ordering and write the expanded
form 
\begin{equation}
U(t,t_{0})=\lim_{\Delta t_{k}\rightarrow 0}[\exp \{-i\tilde{H}
_{N-1}t_{N-1}\}...\exp \{-i\tilde{H}_{0}t_{0}\}]
\end{equation}
At this point we may invoke the assumptions of the BB operations that they
be short, strong pulses and the evolution in between them be that of the
free system-bath.

From this one may also see the connection with the standard control theory
that often uses the final state as the \textquotedblleft
output\textquotedblright\ of the control. This may be used for numerical
algorithms which are associated with a learning or realtime feedback loop.
However, it is clearly desirable to have both the operator and state
pictures (i.e., with and without the explicit state dependence), since one
may often wish to consider the control of the evolution rather than that of
the state.  In fact, in quantum computing, the control objective 
is noiseless evolution rather than simply obtaining a target state.

To optimize the BB procedure, the difference between the BB modified
controls and the ideal evolution should be minimized. This may be achieved
in the continuum by solving the variational problem with a variable
end point. The appropriate variational problem can be formulated as the
minimization of a \emph{cost function} $J$ \cite{Peirce:88,Kosloff:89,Judson:92,Bardeen:97,Rabitz:00,Brif:01}, expressed
in terms of the control fields $\{u_{i}\}$ and cycle time $T_{c}$ as: 
\begin{equation}
J=\int_{t_{0}}^{MT_{c}}\left( \mathrm{Tr}\{[S(u(\tau ),\tau )-S_{w}(\tau )]^{2}\}\right) ^{1/2}d\tau ,
\end{equation}
where $a^{2}\equiv a^{\dagger }a$ and we have used $M$ cycle times $T_{c}$
for the end point (which is not fixed). One may add experimental constraints,
such as finite pulse energy, smoothness of the pulse shapes, etc. \cite
{Peirce:88,Kosloff:89}. This is a standard variational problem for which we would seek $
\delta J=0$ and $\delta ^{2}J<0$. The outcome, i.e., the solution to the
variational problem, will be the optimal control fields $\{u_{i}\}$. Note
that these fields will be approximately continuous for large $M$ and small $
\Delta t$. Then we may approximate them by a discrete set of BB operations
(traditionally defined as piecewise continuous controls; see, e.g., \cite{Jurdjevic:72,Brockett:72}). However, it is to be expected that one of the
advantages of the optimization procedure is that it will yield pulses
that are
easier to implement physically than the pulses coming out of a standard BB
analysis, since the optimization procedure can be formulated to explicitly
take into account experimental constraints. In fact, experience in quantum
chemistry shows that the pulses found by an optimization procedure are often
highly non-intuitive \cite{Peirce:88,Kosloff:89,Judson:92,Bardeen:97,Rabitz:00,Brif:01}.

Questions of convergence, etc., can be avoided by the use of a small number
of cycle times. This will reduce the problem, under the BB assumptions, to a
search on a discrete space. This is the space of discrete, or finite order,
subgroups of unitary groups. Fortunately, for quantum computation, we
require only one- and two-qubit operations which reduces our search spaces
to those of the discrete subgroups of $SU(2)$ and/or $SU(4)$. These have
recently been classified (see \cite
{Fairbanks/Fulton/Klink,Anselmi/etal,Hanany/He} and references therein). We
will not pursue the variational formulation further here. An actual
variational optimization calculation will be presented in a future
publication.


\subsection{Learning Algorithm}

\label{algorithm}

\begin{center}
\begin{figure}[tbp]
\mbox{\epsfig{file=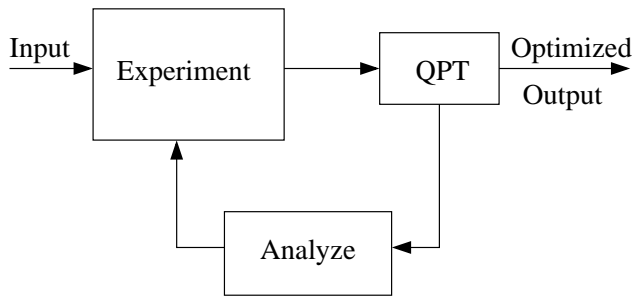}}
\caption{The learning loop control diagram for TQEC.}
\label{loopfig}
\end{figure}
\end{center}

In certain cases it may be possible to perform a large number of experiments
on identically prepared samples, differing in the applied control fields. In
this case, instead of solving a variational problem to find optimal BB\
pulses, one can try to let the experiment guide an off-line learning
algorithm (typically a genetic algorithm) to an optimal solution \cite{Judson:92}.
This algorithm is part of a learning loop, described in Fig.~\ref{loopfig}.

The learning loop consists of the following steps, which are repeated
iteratively in the learning process.

\begin{enumerate}
\item A quantum state is {\texttt{input}} for a particular information
processing task.

\item The state is allowed to interact with a bath and undergo noisy evolution in the {
\texttt{experiment} } Here we may choose to apply BB pulses to modify the evolution. \label{expt}

\item The resulting evolution is obtained through quantum process
tomography, {\texttt{QPT}}.\label{QPT:step}

\item The QPT data is {\texttt{analyzed}} by the learning algorithm to find
an improved BB strategy. This involves solving the key Eqs.~(\ref{eq:one-qubitgates}),(\ref{eq:2qubitcomp}).

\item The previous steps are repeated until convergence to within a
prescribed tolerance.
\end{enumerate}

The result of the procedure is an optimized set of BB\ pulses. This set
includes (i) the least number of BB operations that will reduce or eliminate
the noise in the system, and (ii) the optimal ordering for this minimal set.

Let us emphasize that: (1) no knowledge of the total Hamiltonian or noise
process is assumed (i.e., the optimal implementation is determined
empirically), and (2) no assumption is made about the quality of the BB
operations, only that they should improve the fidelity of the desired
operations. (Of course we know from earlier work \cite{Viola:98,Duan:98e,Vitali:99,Viola:99,Viola:99a,Viola:00a,Zanardi:98b,Zanardi:99d}
that the BB operations should be implemented as strong fast pulses, but
imperfect implementation will still reduce noise.)

Finally, let us note that the learning process could in principle be
incorporated in a real-time feedback loop (e.g., \cite{Doherty/Jacobs/Jungman} and references therein), but this would require a
very fast numerical algorithm to solve Eqs.~(\ref{eq:one-qubitgates}),(\ref{eq:2qubitcomp}).


\section{Concluding Remarks}
\label{ccs}

In order for methods that reduce decoherence and noise in quantum
information processing tasks to succeed in the real world, they must
be confronted with experimental data, and allowed to be optimized in
response to this data. This is the approach we have taken here, in the
context of the dynamical symmetrization, or ``bang-bang'' (BB)
method. We have developed a formulation of the BB method that allows
one to \emph{tailor} the BB control pulses in response to data
acquired by a quantum process tomography experiment. The experiment
supplies a set of numbers that characterize the noise processes
occurring on a short time-scale. From these numbers one can determine
an optimal set of BB pulses, by solving a set of linear equations, in
particular
Eqs.~(\ref{eq:s-sw}),(\ref{eq:main1}),(\ref{eq:one-qubitgates}),(\ref{eq:2qubitcomp}).
These equations correspond to different tasks one may wish to
implement with the help of the BB pulses (respectively, general
storage, single-qubit storage,  single-computation, two-qubit
computation), and yield a set of rotation matrices ($R$) that
correspond to BB pulses that perform the desired tasks.

A promising generalization of a single-shot tomography-BB experiment
is to introduce an off-line learning loop, that uses the above
equations in order to determine an optimized set of BB pulses. The
learning process incorporates tomography measurements from a previous
round in order to find improved BB pulses for the next round. We have
briefly discussed how such a loop, and a concomitant variational
optimization procedure, can be designed.

Throughout
this work we have emphasized that our results have an intuitive
interpretation in terms of a geometric picture, wherein the effect of
BB pulses is to rotate a coordinate vector representing a noisy
Hamiltonian to a desired Hamiltonian.  The geometric picture, via 
Eq.~(\ref{eq:encoded}), also enables the determination of the ability 
to create an encoding (such as a decoherence-free subspace) using empirical data and the available set of 
BB pulses.  Alternatively, it can be used to describe 
the appropriate set of BB operations required to eliminate noise 
from an encoded set of qubits.

We hope that the results presented here will stimulate experiments in
which real data will drive the determination and application of
appropriately tailored BB pulses.


\begin{acknowledgments}
We thank Prof. David Tannor and Prof. Herschel Rabitz for very
useful discussions on optimal control theory. This material is based on
research sponsored by the Defense Advanced Research Projects Agency under
the QuIST program and managed by the Air Force Research Laboratory (AFOSR),
under agreement F49620-01-1-0468 (to D.A.L.). The U.S. Government is authorized to
reproduce and distribute reprints for Governmental purposes notwithstanding
any copyright notation thereon. The views and conclusions contained herein
are those of the authors and should not be interpreted as necessarily
representing the official policies or endorsements, either expressed or
implied, of the Air Force Research Laboratory or the U.S. Government.
\end{acknowledgments}



\end{document}